\documentclass[reprint, pra,floatfix]{revtex4-1}
\usepackage{graphicx}
\usepackage{amssymb}
\usepackage{amsmath}
\usepackage{hyperref}
\usepackage{color}
\begin{document}
\title{Quantum sensors based on weak-value amplification cannot overcome decoherence}
\author{George~C.~Knee}
\email{george.knee@materials.ox.ac.uk}
\affiliation{Department of Materials, University of Oxford, Oxford OX1 3PH, United Kingdom}
\author{G.~Andrew~D.~Briggs}
\affiliation{Department of Materials, University of Oxford, Oxford OX1 3PH, United Kingdom}
\author{Simon~C.~Benjamin}
\affiliation{Department of Materials, University of Oxford, Oxford OX1 3PH, United Kingdom}
\affiliation{Centre for Quantum Technologies, National University of Singapore, 3 Science Drive 2, Singapore 117543}
\author{Erik~M.~Gauger}
\email{erik.gauger@nus.edu.sg}
\affiliation{Centre for Quantum Technologies, National University of Singapore, 3 Science Drive 2, Singapore 117543}
\affiliation{Department of Materials, University of Oxford, Oxford OX1 3PH, United Kingdom}
\date{\today}    
\begin{abstract}
Sensors that harness exclusively quantum phenomena (such as entanglement) can achieve superior performance compared to those employing only classical principles. Recently, a technique based on postselected, weakly-performed measurements has emerged as a method of overcoming technical noise in the detection and estimation of small interaction parameters, particularly in optical systems. The question of which other types of noise may be combated remains open. We here analyze whether the effect can overcome decoherence in a typical field sensing scenario. Benchmarking a weak, postselected measurement strategy against a strong, direct strategy we conclude that no advantage is achievable, and that even a small amount of decoherence proves catastrophic to the weak-value amplification technique. 
\end{abstract}
\maketitle
\section{Introduction}
The prospect of enhancing the sensitivity of nanoscale detectors has recently emerged in light of a greater understanding of fundamental aspects of quantum theory~\cite{GiovannettLloydMaccone2004}. In particular, the appropriate application of entangled states can offer sensing performance at the Heisenberg limit: far superior to the standard quantum limit imposed on separable states~\cite{GiovannettLloydMaccone2004,PezzeSmerzi2009, JonesKarlenFitzsimons2009,ZwierzPerez-DelKok2012}. Another feature of quantum physics not reproducible classically is an effect known as weak-value amplification: in this work we analyze its impact on the field of quantum metrology. 

In a seminal paper of 1988, Aharonov, Albert and Vaidman (AAV) presented a curious quantum mechanical thought experiment giving rise to a quantity  the authors called `the {\it weak value} of a quantum variable'~\cite{AharonovAlbertVaidman1988}. AAV defined the quantity $$A_w:=\langle \psi_f|\textbf{A}|\psi_i\rangle/\langle\psi_f|\psi_i\rangle,$$ generalizing the usual expectation value $\langle A \rangle=\langle \psi_i|\mathbf{A}|\psi_i\rangle$. 
Obtaining $A_w$ involves  i) initializing a quantum system of interest (henceforth `system') into the state $|\psi_i\rangle$; ii) coupling the system weakly to an ancillary measuring device  (henceforth `meter') through the system operator $\textbf{A}$; let the coupling strength be parameterized through a constant $G \in [0,1]$; iii) post-selecting the system into a definite final state $|\psi_f\rangle$. The meter can then be interrogated at full strength to reveal something about the system. A surprising interference effect arises when postselection and weak measurement are combined; i.e. $G < 1$ is chosen such that the deflection induced in the meter is much less than its inherent uncertainty, and when the pre- and postselected states are close to orthogonal, $\langle \psi_f|\psi_i\rangle\rightarrow 0$, one may extract weak values $A_w$ much larger than $\textrm{max}(\langle A\rangle)$ through appropriate measurements of the meter. AAV's expression for $A_w$ has a limited range of validity~\cite{DuckStevensonSudarshan1989}: however, exact treatments reveal that the qualitative effect persists outside this range \cite{WuLi2011,KofmanAshhabNori2012,NakamuraNishizawaFujimoto2012}.  Weak values have been obtained experimentally, see e.g.~Refs.~\cite{RitchieStoryHulet1991,PrydeOBrienWhite2005}. 

In this Article, we shall concentrate on technological applications inspired by AAV's work: the use of the larger-than-usual deflection of the meter for increasing the sensitivity of a suitably designed detector. Although the idea of weak-value amplification (WVA) is mentioned in Ref.~\cite{AharonovAlbertVaidman1988}, its true utility has only recently begun to transpire. AAV considered an amplification of the deflection imparted to a beam of spin-1/2 particles passing a Stern-Gerlach magnetic field gradient. A year later, Duck {\it et al.} described an analogous experiment involving the displacement of a coherent light beam by a birefringent crystal~\cite{DuckStevensonSudarshan1989}. In both cases the internal state of each element of the beam ensemble is preselected, coupled weakly to its spatial wavefunction and finally postselected; leaving the meter (the element's spatial wavefunction) in a characteristically broad distribution peaked around $A_w$. Ordinarily (without postselection), the shift in the meter wavefunction could well be too slight to detect, due to the finite spatial resolution or misalignment of the detection apparatus. Such imperfections can be thought of as classical randomness occurring after the wave function has collapsed, and are known as examples of \emph{technical noise}. This type of noise does not affect the quantum state prior to measurement, but obscures the results after measurement. It can be mitigated by the increased deflection achieved by postselecting the beam into an unlikely internal state. 

In 2008, a tiny lateral displacement of a light beam was promoted to a detectable shift in this way, revealing for the first time the spin Hall effect of light by improving the signal's ratio to technical noise by a factor of $10^4$~\cite{HostenKwiat2008}. In another experiment, a similar amplification was used to boost the angular deflection of light in a Sagnac interferometer~\cite{DixonStarlingJordan2009}. These examples suggest that, if one is limited by technical noise (such as finite detector resolution or imperfect initialization of the meter~\cite{Kedem2012}), the weak-value interference phenomenon can be a useful tool to detect tiny shifts which are  hard to resolve by other means.

Experimental setups are typically afflicted by varying types and magnitudes of noise. If many types of noise are present, one is motivated to overcome the noise that is acting as a limiting factor. Whilst a larger shift in the meter wavefunction is clearly beneficial when technical noise is the dominant source of error, other types of noise may not be overcome so easily with the WVA technique. Shot noise, the intrinsic uncertainty associated with pure state quantum mechanics, may not be defeated~\cite{ZhuZhangPang2011}, although it has been shown that the weak-value technique can match the sensitivity of direct sensing strategies when shot noise dominates, bringing the aforementioned benefits (viz. suppression of technical noise), as it were, `at no extra cost'~\cite{StarlingDixonJordan2009}. What has remained unexplored is whether this remains true when a third type of noise (other than shot noise and technical noise) is considered. Decoherence is a type of noise associated with mixed state quantum mechanics that arises when a system interacts with an unknown environment~\cite{Schlosshau2005}. It randomly affects the dynamics of the quantum state prior to measurement and imposes limits on the sensitivity of parameter estimation schemes~\cite{ShajiCaves2007,EscherMatos-FilDavidovic2011}. Can the WVA effect inspire new detectors which resolve small parameter shifts more quickly than standard techniques, thereby defeating decoherence?

The approach taken in this Article is different to the preceding examples in the following sense: instead of investigating and amplifying a preexisting weak interaction between two quantum degrees of freedom (the process of estimating \emph{an interaction parameter}~\cite{HofmannGogginAlmeida2012}), we are interested in the amplification of an arbitrary phase shift caused by a classical field on a first quantum system by coupling it (with arbitrary strength) to a second. We call this the process of estimating \emph{a phase parameter}. The decoherence of the first system then presents the dominant noise in the problem. We do not consider technical noise, allowing ourselves ideal measurement and control of the quantum system but not its environment. In assessing the performance of WVA, we shall not be satisfied merely with anomalously large expectation values (as intimated by the weak value), but rather demand that the particular statistical model associated with it exhibits a \emph{higher informational content} than is otherwise possible.

The remainder of this Article is organized as follows: in Sections II and III we introduce a canonical phase estimation problem and apply the Fisher information functional to ascertain the performance of what we call the `direct strategy'. In Section IV we explore the possibility of using weak measurements alone, combining this in Section V with postselection, allowing us to evaluate the performance of the `WVA  strategy'. We summarize our conclusions in Section VI. 
\section{Spin based sensing of magnetic field}
The paradigmatic example of a phase parameter estimation problem is a spin-$1/2$ particle in a strong magnetic field $B$ along the $z$-direction. The strong magnetic field leads to a Zeeman splitting of the spin's energy levels, allowing it to be modelled as a quantum two level system commonly referred to as a `qubit'. If the Bloch vector describing the quantum state of the qubit is initialized in the equatorial plane (i.e. into a state unbiased with respect to the `up' $|0\rangle$ and `down' $|1\rangle$ eigenstates) it will precess around the $z$ axis at a rate proportional to the strength of $B$. Moving into a frame that rotates with this evolution, small deviations $\delta B$ in the magnetic field strength are seen as slower evolutions of the Bloch vector around the $z$-axis. 
In this frame, and in the computational $\{|0\rangle,|1\rangle\}$ basis, the evolution of the spin is described by
\begin{eqnarray}
2\rho_{11}(t)&=&1  \nonumber\\
2\rho_{12}(t)&=&-i\textrm{e}^{-i g \delta B t}\Xi(t);
\label{rho}
\end{eqnarray}
$\rho$ is the 2-dimensional density matrix of the spin, and $g$ is its gyromagnetic ratio which we set to unity without loss of generality. As usual, completeness demands $\rho_{22}=1-\rho_{11}$ and $\rho_{21}=\rho_{12}^*$. 
We take $\Xi(t)\in [0,1]$ to be a real, non-negative function (assumed to be independent of $\delta B$) which describes the attenuation of the off-diagonal terms. By choosing this function appropriately one can model, for example, pure dephasing ($\Xi(t)=\textrm{e}^{-\Gamma t}$) or 1/f noise type decay ($\Xi(t)=\textrm{e}^{-\Gamma t^2}$) with a characteristic rate $\Gamma $~\cite{Burkard2009}. The dependence on $t$ is important but sometimes suppressed in our notation. $\delta B$ is an unknown quantity and is the subject of the parameter estimation problem. Whilst for clarity we describe the tangible example of field sensing with a spin, our results apply to many other phase estimation scenarios.

The density matrix, through exposure to the magnetic field, acquires an increasing amount of information about $\delta B$ as time progresses. However, the build-up of useful information is in competition with injurious dephasing mechanisms, which invariably wash out all information after a sufficiently long time. To estimate the value of the parameter $\delta B$, one exploits the causal influence of $\delta B $ on the statistics of outcomes when measuring the spin after an appropriately chosen exposure time. Statistical inferences about $\delta B$ may then be made if one has access to many identical preparations of the density matrix.
\section{Quantum Fisher Information}
Previous studies have commonly assessed the WVA approach (for the purpose of estimating an interaction parameter in the presence of technical noise) using the signal to noise ratio, or the ratio of meter deflection to detector resolution~\cite{ParksGray2011,AshhabYouNori2009,NishizawaNakamuraFujimoto2012,BrunnerSimon2010,FeizpourXingSteinberg2011,Kedem2012,DixonStarlingJordan2009,StarlingDixonJordan2009}. 
By contrast, the Fisher information~\cite{BraunsteinCaves1994,Helstrom1969,Kholevo1982,Fisher1925} is used in parameter estimation when one does not wish to make assumptions about specific measurement limitations;  it can be thought of as a measure of how much information about a given parameter (in this case $\delta B$) is obtainable from a particular \emph{statistical model}: the family of probability distributions generated by the density matrix $\rho$ when it is measured. In the context of weak-value amplification the Fisher information was previously employed by Ref.~\cite{HofmannGogginAlmeida2012}, and Hofmann has argued for a formal connection between the Fisher information and AAV's weak value~\cite{Hofmann2011}.
The Fisher information is defined as 
 \[
F:=\sum_k \frac{1}{p(k|\delta B)}\left(\partial_{\delta B} p(k|\delta B)\right)^2,
\]
where $\partial_{\delta B}$ denotes the partial derivative with respect to the parameter that is to be estimated ($\delta B$) and $p(k|\delta B)$ is the conditional probability of getting outcome $k$ given the value of $\delta B$. The parameter indexes a continuum of differing probability distributions over measurement outcomes: the outcomes in this context being e.g. `spin up' or `spin down' with the probabilities given by the generalized Born rule
\[
p(k|\delta B)=Tr[\rho\Pi_k],
\]
where $\Pi_k$ is the POVM element associated with outcome $k$.  The maximum likelihood estimation procedure involves examining these probabilities and guessing a value of $\delta B$ that would generate the observed frequencies with the greatest probability, inverting the statistics in a Bayesian sense~\cite{Myung2003}. The Fisher information is then inversely proportional to the variance in the estimate of $\delta B$, and provides a good indication of the performance of any given statistical model. It is clear that the choice of measurement (POVM) will affect the Fisher information. 

Fixing the POVM to a sharp measurement in the $\sigma_x:=|0\rangle\langle1|+|1\rangle\langle0|$ basis, one finds for a spin in a field
\begin{equation}
F_{\textrm{d}}=\frac{t^2 \cos^2(t \delta B)\Xi^2}{1-\Xi^2\sin^2(t \delta B )}.
\end{equation}
Here, the subscript `d' is used to denote quantities pertaining to a direct sensing of $\delta B$, i.e. through the density matrix described by Eq.~(\ref{rho}). This expression exhibits oscillations over time as the angle between the quantum state and the measurement basis varies between pessimal (when the measurement basis is parallel to the final state, $ t\delta B =[n+1/2]\pi$) to optimal (when the measurement basis is perpendicular to the final state, $t\delta B =n\pi$).

To eliminate the dependence on the measurement choice one can deploy the optimal POVM: since the estimation procedure involves many samples of the probability distribution, some of them may be used to adaptively update the POVM after an initial guess, causing it to rapidly converge on an optimum~\cite{BrivioCialdiVezzoli2010}. A quantity that captures the maximum $F$ in a variation over all POVMs is the quantum Fisher information (QFI), defined as 
\begin{equation}
H:=2\sum_{nm} \frac{|\langle m|\partial_{\delta B}\rho|n\rangle|^2}{p_n + p_m}.
\label{qfiH}
\end{equation}
The above sum only includes terms for which $p_n+p_m\neq0$ and where $n,m$ index the basis states in the spectral decomposition of $\rho$, $\rho=\sum_i p_i|i\rangle\langle i |$~\cite{Paris2009}. 
In our case
\begin{equation}
H_{\textrm{d}}=\Xi^2 t^2,
\end{equation}
which is, notably, independent of the parameter $\delta B$. The oscillations in time have disappeared, and this leaves only an envelope with turning points which can be found by solving $\dot{\Xi}t^2=-\Xi$. For example when $\Xi=\textrm{e}^{-\Gamma  t}$ the maximum of $H$ occurs at $t^*=1/\Gamma $.

According to the Cramer-Rao bound~\cite{Cramer1999}, the minimum variance of the parameter $\delta B$ is given by $1/ (NH)$ where $N$ is the number of trials. A larger value of $H$ thus entails a smaller minimum variance, which is obtained efficiently through maximum likelihood estimation~\cite{Fisher1925}. The canonical measure of the utility of a detector is the measurement sensitivity $S$, which is the minimum uncertainty achievable in the parameter in a fixed amount of sensing time, allowing for the estimation procedure to be stopped (e.g. after $t^*$), reset and repeated an arbitrary number of times within the fixed duration.  Novel approaches to quantum metrology must show an improvement in $S$ (which depends on $H$) to claim an advantage over established techniques. The aforementioned entanglement enhanced sensors have been shown to have a lower (read superior) sensitivity~\cite{GiovannettLloydMaccone2004,PezzeSmerzi2009, JonesKarlenFitzsimons2009, ModiCableWilliamson2011, SchaffryGaugerMorton2010} by virtue of achieving a square root improvement over the standard approach given the same resources.
\section{Arbitrary strength ancilla measurement}
An ancillary spin can be used as part of the measurement process, often bringing some advantage -- for example in the preamplification of a crystal defect based magnetic field sensor~\cite{SchaffryGaugerMorton2011}. Consider an ancillary qubit (the meter) initialized in the $x$-$y$ plane, coupled to the system qubit so that one has control over the joint system. We envisage a measurement operation which effects the following unitary transformation on the system and meter
\begin{equation}
M(G)=\Pi_+\otimes \mathbb{I}+\Pi_-\otimes \textrm{exp}(iG\pi\sigma_z/2).
\end{equation}¥
$\Pi_\pm$ are projectors onto the $\pm1$ eigenstates of a traceless `control observable'  (of the system) that lies in the plane, $\Pi_+-\Pi_-=\cos\Theta\sigma_y+\sin\Theta\sigma_x$. The choice of $\sigma_z := |1\rangle\langle1|-|0\rangle\langle0|$ as a meter observable in the second term is chosen because the ancilla is initialized in plane~\footnote{best results are obtained when the ancilla measurement basis is unbiased with respect to the axis about which the ancilla rotates }. This setup allows the measurement strength to be varied independently of the initial state of the meter, and hence captured by $G$ alone: the measurement is strong when $G=1$ and gets weaker as $G\rightarrow0$.
\begin{figure}
\includegraphics[width=\linewidth]{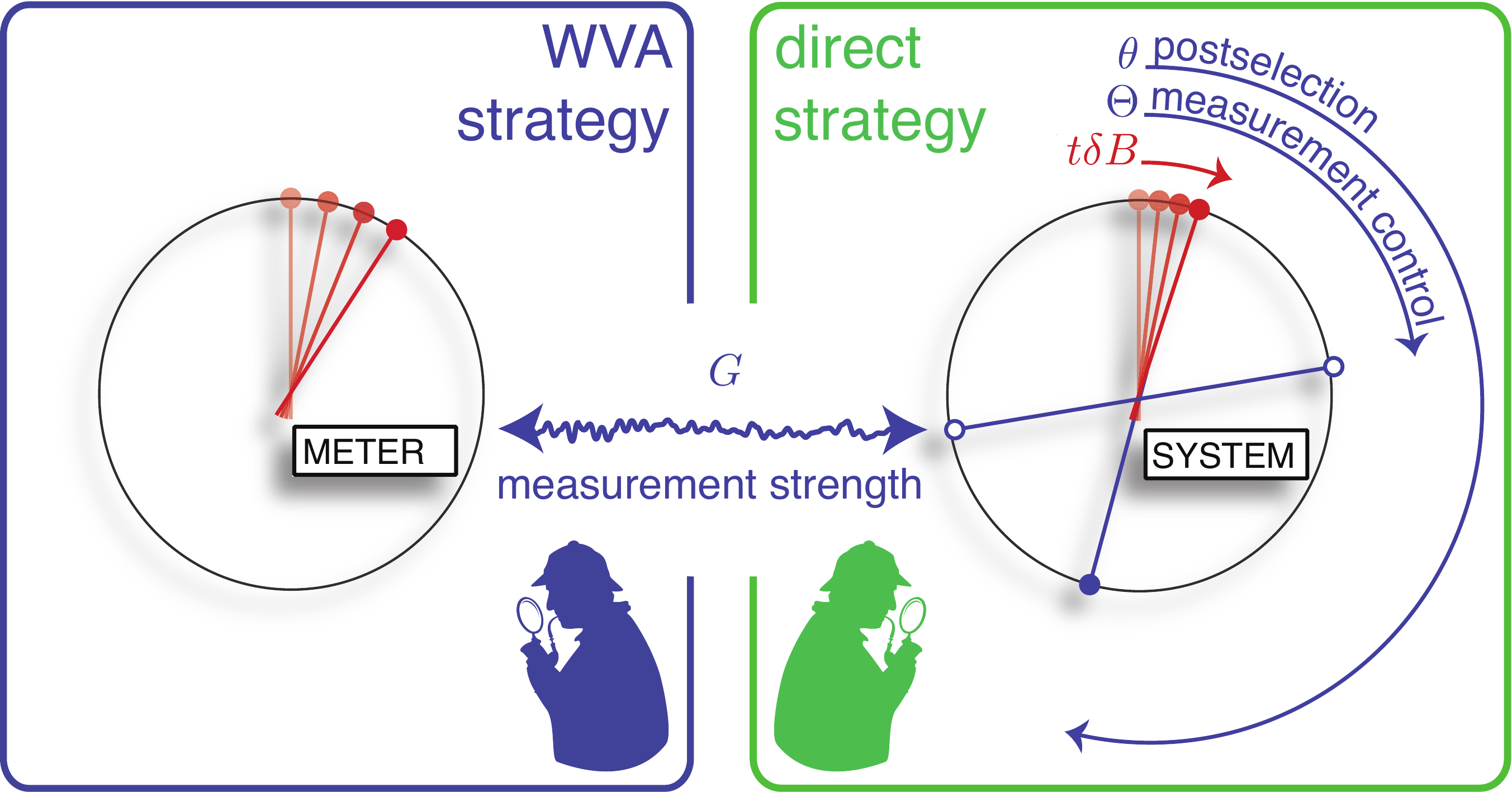}
\caption{\label{schematic}(Color online) A schematic for two different approaches to parameter estimation using spins; the view is into the equatorial plane of the Bloch sphere. A direct strategy relies only on interrogations of the system spin, which picks up a field dependent phase $\textrm{exp}(it\delta B)$ over time.  A WVA  (weak-value amplified) strategy makes use of an ancillary spin in the hope of gaining an advantage. The latter technique involves coupling the first spin with a second `meter' spin with variable strength, and then interrogating the meter spin only if the system spin is successfully postselected into a certain final state. Such a postselection can be achieved by the use of projective measurements in an appropriate basis given by $\theta$, and is known to lead to a larger than expected deflection of the meter spin if the coupling $G$ is weak. Note that whilst $\delta B$ is an unknown quantity of interest, $\theta,\Theta,G$ are tunable by the experimenter in the WVA strategy (their meanings are explained in the main text).}
\end{figure}¥
After tracing out the system spin, one finds that the QFI of the meter spin is 
\begin{equation}
H_{\textrm{anc}}=\frac{\Xi^2t^2 \sin^2(G \pi/2) \sin^2(\Theta -t \delta B) }{1-\Xi^2\cos^2(\Theta -t \delta B) }.
\end{equation}
The subscript `anc' denotes quantities pertaining to an indirect sensing protocol, involving the use of an ancilla spin coupled to the system spin with arbitrary measurement strength. Assuming the experimenter has control over the measurement operation, it is clear that she arranges $\Theta=t\delta B+\pi/2$ for best results (which corresponds to the control observable being unbiased w.r.t.~the system state), whence
\begin{equation}
H_{\textrm{anc}}\rightarrow\Xi^2t^2 \sin^2(G\pi/2).
\end{equation}
This matches the performance of the direct strategy when the measurement corresponds to a full controlled $\pi-$rotation, $G=1$.  The WVA scheme would seem to  sacrifice performance by operating near to $G=0$ (where the dependence of $H_{\textrm{anc}}$ on $G$ is roughly quadratic), however, we shall see that the postselection step of the protocol leads to an amplification which mitigates this apparent loss of information. 
\section{Postselected strategy}
\begin{figure}[t]
\includegraphics[width=\linewidth]{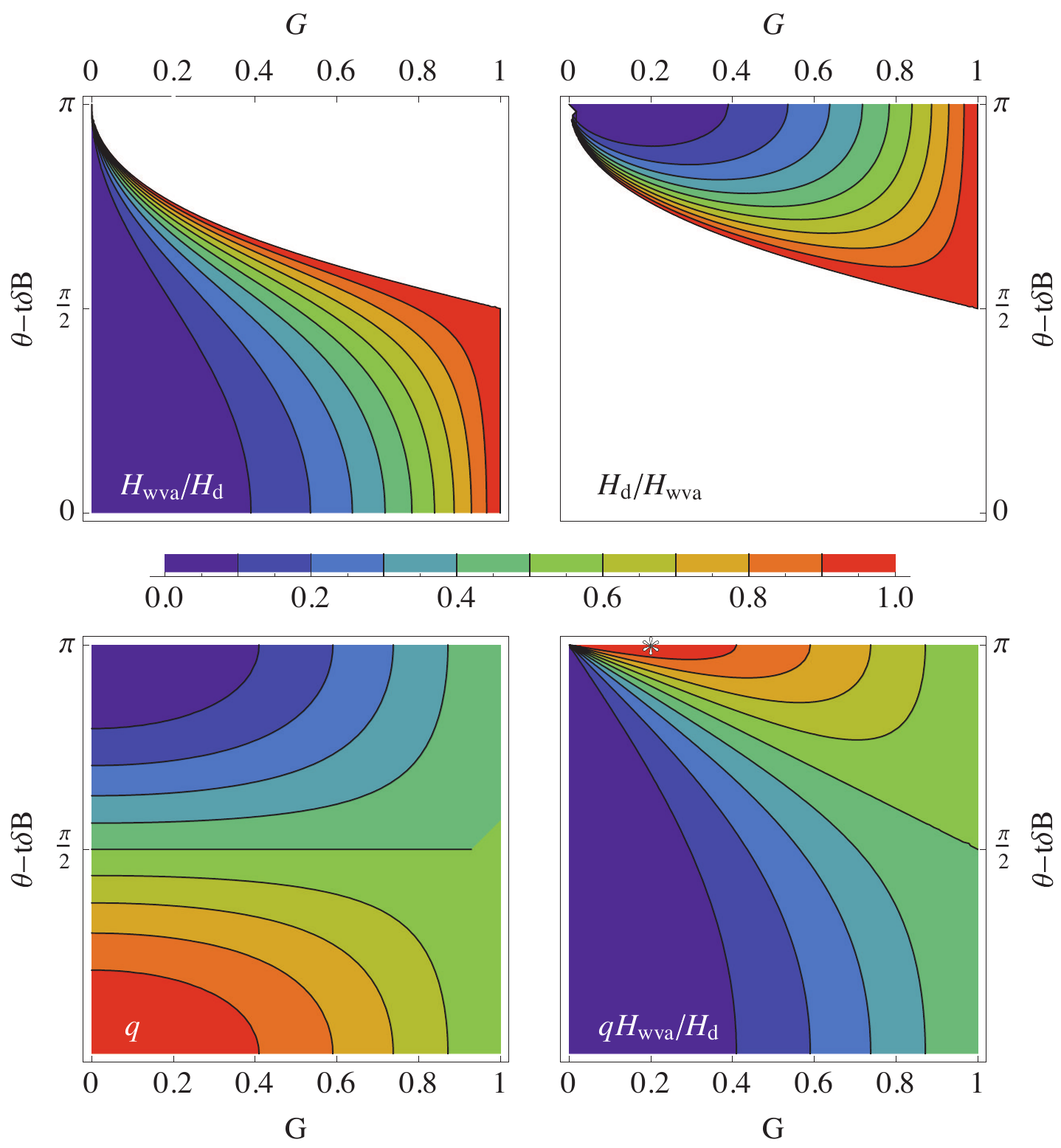}
\caption{\label{contours}(Color online) Contour lines of four important quantities are plotted against measurement strength $G$ and the angle ($\theta-t\delta B$) between pre- and postselection. \emph{UL pane:} $H_{\textrm{wva}} : H_{\textrm{d}}$ uncorrected for the low postselection probability. White regions show a ratio higher than one and hence the apparent superiority of the WVA scheme over a direct scheme; \emph{UR pane:} the inverse ratio, white regions now show where the WVA scheme is inferior; \emph{LL pane:} the postselection success probability $q$; \emph{LR pane:} shows the corrected ratio $qH_{\textrm{wva}}:H_{\textrm{d}}$, note there are no longer white regions and the weak strategy is never better. The $*$ denotes a choice of $(G,\theta)$ that is generalized to include dephasing noise in Fig.~\ref{ratioasfunctionoftandgamma}. }
\end{figure}

Now we introduce postselection into the protocol. The system spin is allowed to pick up phase in the weak field as usual. After a given time the measurement $M$ is triggered: this entangles the system spin with the meter, which is initialized into the $-1$ eigenstate of $\sigma_y$, $\eta(0)=|i^-\rangle\langle i^-|$. The system spin is then measured, and only if it is found in a certain postselection state $|\psi_f\rangle=(|0\rangle+\textrm{e}^{i\theta}|1\rangle)/\sqrt{2}$, the ancillary spin is interrogated in the usual manner (using an adaptive maximum likelihood estimation procedure). 
In the most general case the angle between pre- and postselection can be varied independently of the measurement control angle (as sketched in Fig.~\ref{schematic}).
The density matrix of the meter after the measurement interaction and post-selection will be given by  \footnote{Identity matrices are implied in the following expression, i.e.~${\langle \psi_f|_s=\langle \psi_f|\otimes \mathbb{I}}$ etc.}
\begin{equation}
\eta(t)\propto\langle \psi_f|_sM\left[\rho(t)\otimes\eta(0)\right]M^{\dagger}|\psi_f\rangle_s, \nonumber
\end{equation}
with the proportionality constant fixed by normalization. Under these operations, the evolution of the meter is confined to the $x$-$y$ plane, which is optimal for phase estimation~\cite{TekluGenoniOlivares2010}. The full evolution of $\eta$ is given in Appendix A, but here we specialize, without loss of generality, to $\theta=\Theta+\pi/2$ (corresponding to a postselection state that is unbiased w.r.t.~the control observable \footnote{One can choose instead to have the preselection unbiased w.r.t.~the control observable, as for the $H_{\textrm{anc}}$ above, but one does no better in that case.}). Thus we obtain, after normalizing, the simplified expression:
\begin{widetext}
\begin{align}
\eta_{11}(t) &=\frac{1}{2} \nonumber \\
\eta_{12}(t) &=\eta_{12}(t)=\frac{e^{-\frac{1}{2} i \pi G} \left[\Xi \sin \left(\frac{ G\pi}{2}\right) \sin (\theta - t\delta B )+i \left(\cos \left(\frac{G\pi}{2}\right)+\Xi \cos (\theta - t\delta B )\right)\right]}{2 \Xi \cos \left(\frac{ G\pi}{2}\right) \cos (\theta - t\delta B )+2}. \label{eta}
\end{align}			
\end{widetext}
Of course, $\eta_{22}=1-\eta_{11}$ and $\eta_{21}=\eta_{12}^*$. Calculating the QFI can be difficult for arbitrary density matrices, but we found that applying Eq.~(\ref{qfiH}) to a general qubit state in the equatorial plane leads to 
\begin{equation}
H=\frac{(y^2-1)\hat{x}^2-2xy\hat{x}\hat{y}+(x^2-1)\hat{y}^2}{x^2+y^2-1},
\end{equation}
where $x+iy=2\eta_{12}(t)$ and `$~\hat{ }~$' denotes partial derivative w.r.t.~$\delta B$. One straightforwardly obtains, taking real and imaginary parts of (\ref{eta}), 
\begin{equation}
H_{\textrm{wva}}=H_{\textrm{d}}
 \sin^2(G\pi/2) A.
 \end{equation}
The subscript `wva' denotes the use of an indirect sensing strategy through the density matrix (\ref{eta}); but note that we have treated the measurement strength entirely generally. The expression admits the following interpretation: the first term is the bare information available through direct techniques; the second term represents the cost of having a finite strength measurement, and is present with and without postselection. The final term
\begin{equation}
A=(1+\Xi(t)\cos(G\pi/2)\cos(\theta-t\delta B))^{-2}
\end{equation}
is due to the weak-value amplification effect: it becomes large when both $\theta-t\delta B$ is close to an odd integer multiple of $\pi$ and $\cos(G\pi/2)\approx1$, i.e.~for a weak measurement strength and almost orthogonal pre- and postselection. The upper two panels of Fig.~\ref{contours} demonstrate that in this regime $H_{\textrm{wva}} > H_{\textrm{d}}$, and the WVA strategy would seem to outperform the direct one.
However,  the much reduced postselection probability
\begin{equation*}
q=Tr[(|\psi_f\rangle\langle \psi_f|\otimes \mathbb{I})M\rho(t)M^\dagger]
\end{equation*}
must be taken into account, evaluating to $q=1/(2\sqrt{A})$. Note that $q$ is nonzero when pre- and postselection are orthogonal due to the back action of the weak measurement~\cite{WuLi2011}. Once the probability is properly accounted for,  
\begin{equation}
qH_{\textrm{wva}}=\frac{\Xi^2 t^2 \sin^2(G \pi/2)}{2 (1+\Xi\cos(G \pi/2) \cos(\theta -t \delta B))}.
\label{mainresult}
\end{equation}
A fair comparison between the efficiency of the weak-value approach and the direct approach can be made by considering the ratio of the two strategies. Before any optimization this ratio, shown in the lower right panel of Fig.~\ref{contours},  is given by 
\begin{equation}
\frac{qH_{\textrm{wva}}}{H_{d}}=\frac{\sin^2(G \pi /2)}{2(1+\Xi\cos\left(G \pi /2\right) \cos(\theta -t \delta B ))}.
\end{equation}
Note that by inspection this expression never exceeds unity for \emph{any} function $\Xi(t)\in[0,1]$. Since the inequality $qH_{\textrm{wva}}<H_{\textrm{d}}\quad \forall t$ implies $S_{wva}>S_d$ (where $S$ is the measurement sensitivity), this argument is sufficient to establish that the WVA technique can never reach a better (i.e. lower) sensitivity than a direct technique.

\begin{figure}
\includegraphics[width=\linewidth]{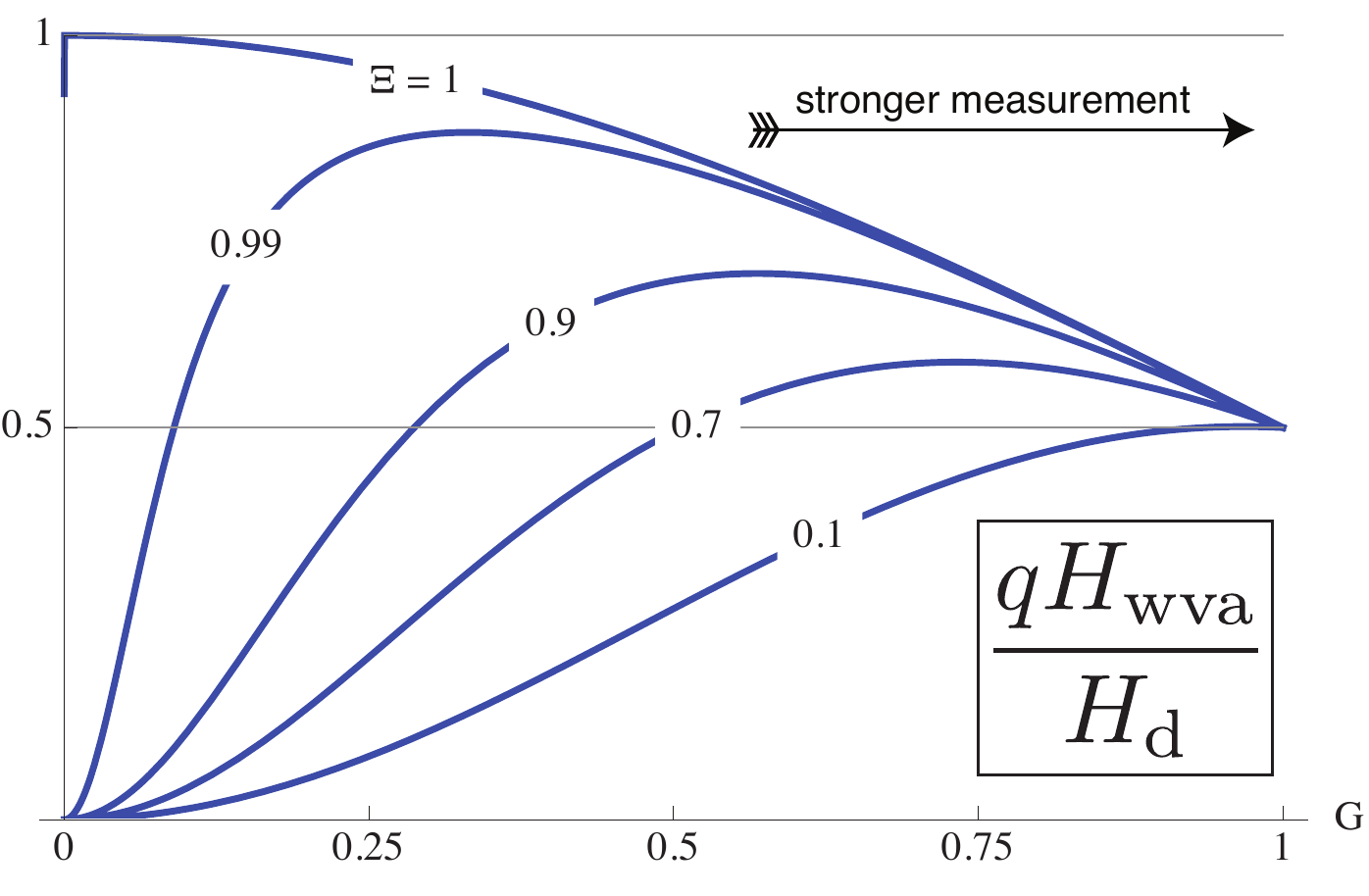}
\caption{\label{ratioasfunctionofG}(Color online) Ratio of information available in the WVA strategy and in its direct counterpart, as a function of the measurement strength $G$. The different curves correspond to various values of the attenuation function $\Xi(t)$. The postselection is fixed to $\theta=t\delta B+\pi$. When $G=1$, one should match the direct strategy, only the postselection probability $q=1/2$ implies by symmetry that half of the information is thrown away (see Appendix B).}
\end{figure}

 Note that when dephasing noise is completely absent then one can reach a ratio of unity for the correct choice of $G$ and $\theta$; this is in good agreement with the results of Starling {\it et al.}~\cite{StarlingDixonJordan2009} and Zhu {\it et al.}~\cite{ZhuZhangPang2011}. However, even a small attenuation is catastrophic to the weak-value technique because for small $G$ there is a faster than exponential 
decay of the ratio $q H_{\textrm{wva}} : H_{\textrm{d}}$ as $\Xi$ decreases (see Fig.~{\ref{ratioasfunctionofG}}), and a strong measurement quickly becomes favorable. 
¥

To illustrate this behavior with a concrete example, the time dependence of $H_{\textrm{d}}$ and $H_{\textrm{wva}}$ is shown in Fig.~\ref{ratioasfunctionoftandgamma}  for phenomenological dephasing noise, $\Xi(t) = e^{- \Gamma t}$. 
In Appendix B we address the possibility of keeping \emph{all} of the data after the postselection measurement. We also generalize the decoherence model to incorporate, e.g. amplitude damping processes, and consider other scenarios where WVA may be of benefit: namely those scenarios where the measurement itself is not implemented cleanly, and when the interaction is unavoidably weak (Appendix D). None of these generalizations alter our conclusion.
\begin{figure}
\includegraphics[width=\linewidth]{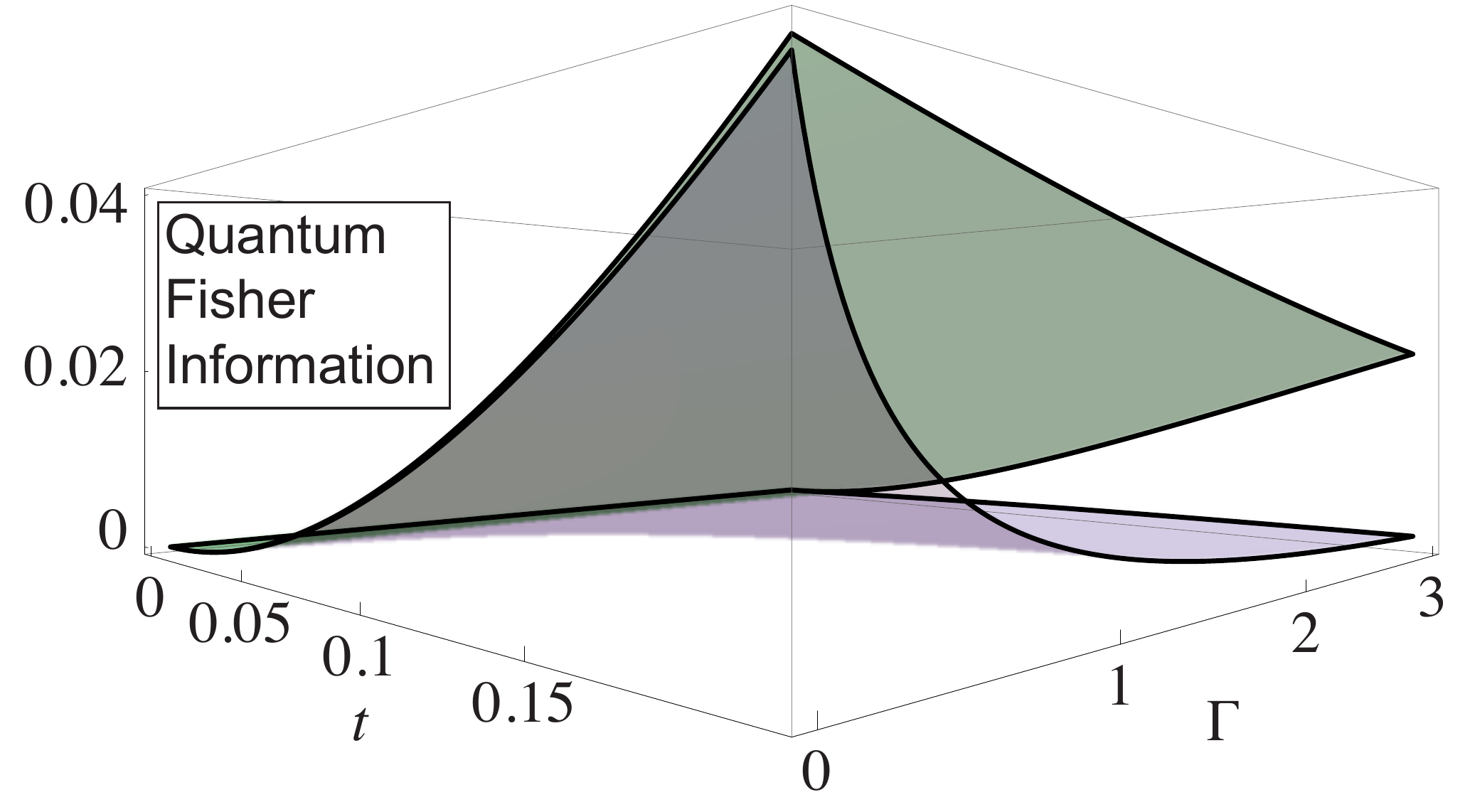}
\caption{\label{ratioasfunctionoftandgamma}(Color online) Quantum Fisher information (in units of $\textrm{T}^{-2}$ ) for the two competing strategies is plotted as a function of time $t$ and the dephasing rate  $\Gamma$ (for $\Xi=\textrm{e}^{-\Gamma t}$). The upper green surface corresponds to the direct strategy $H_\textrm{d}$ and the lower purple surface corresponds to $qH_{\textrm{wva}}$, the corrected weak-value amplified strategy. The postselection and measurement strength are fixed to $\theta=t\delta B+\pi, G=0.02$, respectively, corresponding to the $*$ in Fig.~\ref{contours}. Even a moderate amount of dephasing has a catastrophic effect on the weak-value scheme.}
\end{figure}¥
\section{Conclusion}
We have analyzed the utility of weak-value amplification for the purpose of estimating an unknown phase parameter appearing in the Hamiltonian of a two level quantum system, finding no advantage over strong and direct techniques for the broad class of noise models captured by Eq.~(\ref{rho}). This includes any kind of dephasing noise, and we show in Appendix C that the quality of our conclusion is preserved when other imperfections (such as $T_1$ processes) are considered. 
When decoherence is completely absent the WVA strategy can match the performance of the direct approach, encouraging the motto:  `one postselected run acting as though many unpostselected runs'~\cite{FeizpourXingSteinberg2011}. In contrast to entanglement or discord enhanced sensing protocols, however, which are robust against a degree of mixing of the quantum state~\cite{ModiCableWilliamson2011}, any level of dephasing noise ruins the performance of the WVA approach.

While we have described the system-meter qubit pair as spin-1/2 particles, they are isomorphic to many other physical systems: for example one can use the polarization states of photons to measure a phase shift introduced by a crystal, and couple photons together to enable weak measurement~\cite{PrydeOBrienWhite2005}. 

We reiterate that our results do not contradict studies that have already put WVA to use experimentally~\cite{HostenKwiat2008,DixonStarlingJordan2009,StarlingDixonJordan2009} or the many theoretical proposals for improving signal to noise~\cite{ZilberbergRomitoGefen2011,BrunnerSimon2010,Kedem2012,FeizpourXingSteinberg2011,HofmannGogginAlmeida2012,ShikanoTanaka2011} since in those cases the quantity of interest is an interaction parameter, and only technical noise is overcome. When the limiting disturbance is to the quantum state however, rather than to the classical information following the measurement, then there is no advantage to be gained by using a weak-value amplified approach. 

Future work may elucidate whether imaginary weak values can be more useful for metrology than real weak values, as has been suggested by Refs.~\cite{BrunnerSimon2010,ParksGray2011}; it would be interesting to study how this might apply in finite dimensional meters~\cite{WuMolmer2009}. The issue of technical noise in finite dimensional meters could also be studied. 
\\~\\
\section{acknowledgements}
This work was supported by the Engineering and Physical Sciences Research Council (EPSRC), the John Templeton Foundation, and the National Research Foundation, the Ministry of Education, Singapore. \\~\\
\appendix
\section{Full evolution of ancilla}
For a completely general choice of angles between pre- and postselection and the measurement control as defined in Fig.~\ref{schematic}, the evolution of the meter quit is described by 
\begin{widetext}
\begin{eqnarray}
\eta_{11}=&\frac{1}{2}& \nonumber \\
\eta_{12}=&\frac{4 i e^{-\frac{1}{2} i G \pi } \left(\cos\left(G \pi /2\right)+i \cos(\theta -\Theta ) \sin\left(G \pi /2\right)+\left(\cos\left(G \pi /4\right)^2 \cos(\theta -t \omega )-\cos(\theta -2 \Theta +t \omega ) \sin\left(G \pi /4\right)^2+i \cos(\Theta -t \omega ) \sin\left(G \pi /2\right)\right) \Xi \right)}{8+8 \left(\cos\left(G \pi /4\right)^2 \cos(\theta -t \omega )+\cos(\theta -2 \Theta +t \omega )\sin\left(G \pi /4\right)^2\right) \Xi}.&
\end{eqnarray}
\end{widetext}
In the main text a fixed relationship between $\Theta$ and $\theta$ was chosen for simplicity, and because this additional freedom cannot provide an advantage.
\section{Using all the data}
Since the WVA technique can get close to the performance of a direct strategy and involves discarding the majority of experimental runs, one might imagine that some of the discarded data may be used to increase the information, perhaps even allowing the technique to outperform the direct strategy. In addition to the quantum Fisher information arising from successful postselection Eq.~(\ref{mainresult}), one now has
\begin{equation}
(1-q)H^\perp_{\textrm{wva}}=\frac{t^2\Xi^2 \sin^2\left(G \pi /2\right)}{2-2 \Xi  \cos\left(G \pi /2\right) \cos(\theta -t \omega )}
\end{equation}
resulting from runs that would ordinarily be discarded. 

\begin{figure}
\includegraphics[width=\linewidth]{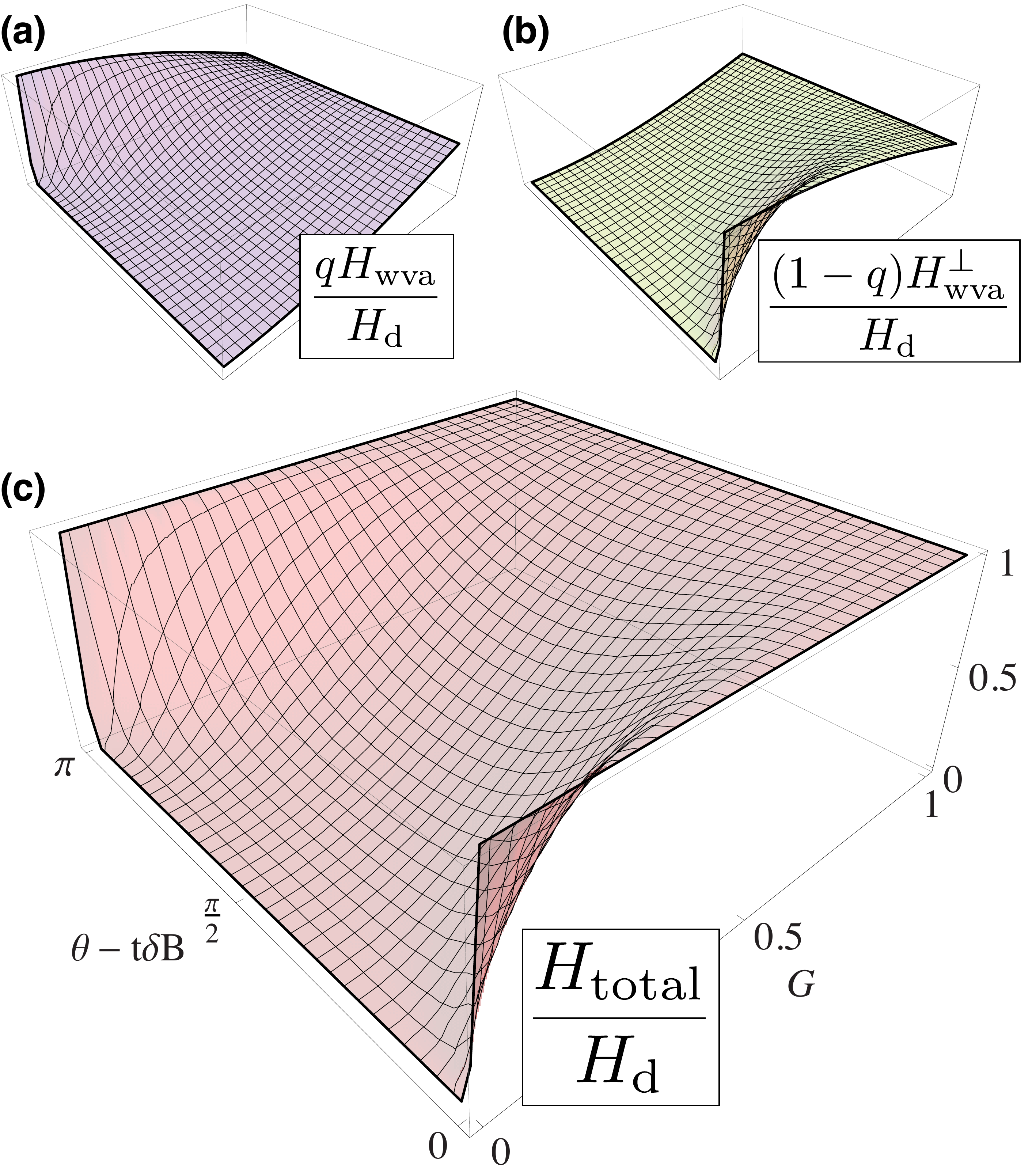}
\caption{\label{total}(Color online) Quantum Fisher information for a weak-value amplified sensing strategy {\bf(a)} when the postselection is `successful' {\bf (b)} when the postselection is `unsuccessful' {\bf (c)} when both are considered together; the plots shown correspond to the decoherence-free case.}
\end{figure}

One can see that the two quantities are complementary in the following sense. In the regime where the WVA effect is strongest, the discarded runs carry less and less information. Since the total information is additive, one can achieve
\begin{eqnarray}
H_{\textrm{total}}:=&qH_{\textrm{wva}}+(1-q)H^\perp_{\textrm{wva}}\\
=&\frac{t^2 \sin^2\left(G \pi /2\right)}{1-\Xi ^2 \cos^2\left(G \pi /2\right) \cos^2(\theta -t \omega)}.
\end{eqnarray}¥
This quantity cannot exceed $H_{\textrm{d}}$, but has some interesting features. It is greater than $qH_{\textrm{wva}}$ in regions where the WVA effect is small, notably reaching  $q H_{\textrm{wva}} / H_{\textrm{d}}=1$ (rather than $1/2$) when the measurement is fully strong. It converges on $qH_{\textrm{wva}}$ when the WVA effect is pronounced, see Fig.~\ref{total}.
¥
\section{A more general noise model}
One can allow the populations of the density matrix $\rho$ to have a time dependence, 
\begin{eqnarray}
2\rho_{11}(t)&=&R(t)  \nonumber\\
2\rho_{12}(t)&=&-i\textrm{e}^{-i g \delta B t}\Xi(t).
\end{eqnarray}
Note that $R(t)$ and $\Xi(t)$ are not independent and are related by $\Xi \leq 2\sqrt{R-R^2}$, and that decoherence models such as relaxation can now be modelled $R(t)=\textrm{e}^{-t/T_1},~\Xi(t)=\textrm{e}^{-t/ (2T_1)}$. This has no effect on $H_{\textrm{d}}$ or $H_{\textrm{wva}}$ (despite the ancilla qubit now having an evolution out of plane that depends on $\delta B$). Notwithstanding further optimizations of the measurement interaction in this scenario, the results presented here are conserved when the noise model is generalized to allow for relaxation, polarization damping or other noise channels. 
\section{Imperfections in the measurement device}
For an exhaustive assessment of the utility of WVA, one can consider more contrived situations. The failure of the technique to provide enhancement over direct measurements on a noisy system is twofold. Firstly it can be thought of as deriving from the order in which the amplification and noise act. Because the noise acts before the measurement takes place, it gets amplified as much as the signal does. Secondly, one can think of the failure as deriving from the drop in signal strength incurred from the weak interaction strength. 
\subsection{Noisy meter system}
It is natural to inquire about the case when the amplification takes place before any noise has acted. Our model can be easily adapted to study this possibility. One imagines a situation where a well-isolated quantum system (perhaps a nuclear spin with an infinite coherence time) is used to sense a parameter (a magnetic field deviation), but is inaccessible to direct measurement. Readout can be performed by coupling to an ancillary spin (perhaps an electron spin) as above, but now the ancilla suffers an attenuation of its coherences described by $\Sigma(t)$. 
An estimation scheme using a strong measurement $G=1$ without postselection can achieve (with the usual optimization over the measurement control)
\begin{equation}
\tilde{H}_{\textrm{anc}}=\Sigma^2t^2.
\end{equation}
The `$~\tilde{ }~$' denotes the decoherence acting on the meter rather than the system. Allowing the measurement strength to vary and enabling postselection yields
\begin{eqnarray}
\tilde{H}_{\textrm{wva}}=&\frac{\Sigma^2t^2\sin^2(G\pi/2)}{(1+\cos(G\pi/2)\cos(\theta-t\delta B))^2}\\
\rightarrow&\Sigma^2t^2\cot^2(G\pi/4)
\end{eqnarray}
when tuning the postselection as above. Unsurprisingly the signatures of AAV's effect persist as long as the attenuation of coherence in the meter has not completely annihilated the off diagonal density matrix terms~\cite{ChoLimRa2010,ShikanoHosoya2010,ThomasRomito2012}. Once more, the weak value approach suggests an emphatic improvement (the expression diverges as $G\rightarrow 0$). Note, however, that assuming the postselection can be implemented is not entirely consistent with the idea of being forced to measure indirectly through a noisy qubit. Nevertheless, with the postselection probability taken into account one finds
\begin{equation}
q\tilde{H}_{\textrm{wva}}=\Sigma^2t^2\cos^2(G\pi/4).
\end{equation}
In close resemblance to the above case, the QFI never exceeds the direct strategy benchmark; in fact, it varies between at best being equal to and at worst half as large. So, even when given this improbably favorable scenario, the WVA approach fails to offer an advantage. 
\subsection{Limited measurement strength}
Let us consider the advantage postselection alone can have on an arbitrary strength measurement accomplished via a meter qubit. Perhaps one has no control over the value of $G$, but is in fact forced to make a weak measurement. In this case the appropriate ratio is
\begin{equation}
\frac{qH_{\textrm{wva}}}{H_{\textrm{anc}}}=\frac{1}{2 \left(1+\Xi\cos\left(\frac{G \pi }{2}\right) \cos(\theta -t \delta B )\right)}.
\end{equation}
Now, there is a clear advantage as long as the dephasing noise is not too aggressive. But again one has had to assume the use of strong measurements for the purpose of postselection but deny their use in the estimation part of the protocol, which is not entirely consistent. This situation may appear similar to the case of estimating an interaction parameter: but there the unknown quantity of interest is the measurement strength, and the Fisher information must be calculated w.r.t.~$G$~\cite{Steinberg2010}. 
\bibliographystyle{apsrev4-1}
\bibliography{gck_full_bibliography}

\begin{thebibliography}{49}%
\makeatletter
\providecommand \@ifxundefined [1]{%
 \@ifx{#1\undefined}
}%
\providecommand \@ifnum [1]{%
 \ifnum #1\expandafter \@firstoftwo
 \else \expandafter \@secondoftwo
 \fi
}%
\providecommand \@ifx [1]{%
 \ifx #1\expandafter \@firstoftwo
 \else \expandafter \@secondoftwo
 \fi
}%
\providecommand \natexlab [1]{#1}%
\providecommand \enquote  [1]{``#1''}%
\providecommand \bibnamefont  [1]{#1}%
\providecommand \bibfnamefont [1]{#1}%
\providecommand \citenamefont [1]{#1}%
\providecommand \href@noop [0]{\@secondoftwo}%
\providecommand \href [0]{\begingroup \@sanitize@url \@href}%
\providecommand \@href[1]{\@@startlink{#1}\@@href}%
\providecommand \@@href[1]{\endgroup#1\@@endlink}%
\providecommand \@sanitize@url [0]{\catcode `\\12\catcode `\$12\catcode
  `\&12\catcode `\#12\catcode `\^12\catcode `\_12\catcode `\%12\relax}%
\providecommand \@@startlink[1]{}%
\providecommand \@@endlink[0]{}%
\providecommand \url  [0]{\begingroup\@sanitize@url \@url }%
\providecommand \@url [1]{\endgroup\@href {#1}{\urlprefix }}%
\providecommand \urlprefix  [0]{URL }%
\providecommand \Eprint [0]{\href }%
\providecommand \doibase [0]{http://dx.doi.org/}%
\providecommand \selectlanguage [0]{\@gobble}%
\providecommand \bibinfo  [0]{\@secondoftwo}%
\providecommand \bibfield  [0]{\@secondoftwo}%
\providecommand \translation [1]{[#1]}%
\providecommand \BibitemOpen [0]{}%
\providecommand \bibitemStop [0]{}%
\providecommand \bibitemNoStop [0]{.\EOS\space}%
\providecommand \EOS [0]{\spacefactor3000\relax}%
\providecommand \BibitemShut  [1]{\csname bibitem#1\endcsname}%
\let\auto@bib@innerbib\@empty
\bibitem [{\citenamefont {Giovannetti}\ \emph {et~al.}(2004)\citenamefont
  {Giovannetti}, \citenamefont {Lloyd},\ and\ \citenamefont
  {Maccone}}]{GiovannettLloydMaccone2004}%
  \BibitemOpen
  \bibfield  {author} {\bibinfo {author} {\bibfnamefont {V.}~\bibnamefont
  {Giovannetti}}, \bibinfo {author} {\bibfnamefont {S.}~\bibnamefont {Lloyd}},
  \ and\ \bibinfo {author} {\bibfnamefont {L.}~\bibnamefont {Maccone}},\
  }\href@noop {} {\bibfield  {journal} {\bibinfo  {journal} {Science}\ }\textbf
  {\bibinfo {volume} {306}},\ \bibinfo {pages} {1330} (\bibinfo {year}
  {2004})}\BibitemShut {NoStop}%
\bibitem [{\citenamefont {Pezz\'e}\ and\ \citenamefont
  {Smerzi}(2009)}]{PezzeSmerzi2009}%
  \BibitemOpen
  \bibfield  {author} {\bibinfo {author} {\bibfnamefont {L.}~\bibnamefont
  {Pezz\'e}}\ and\ \bibinfo {author} {\bibfnamefont {A.}~\bibnamefont
  {Smerzi}},\ }\href@noop {} {\bibfield  {journal} {\bibinfo  {journal} {Phys.
  Rev. Lett.}\ }\textbf {\bibinfo {volume} {102}},\ \bibinfo {pages} {100401}
  (\bibinfo {year} {2009})}\BibitemShut {NoStop}%
\bibitem [{\citenamefont {Jones}\ \emph {et~al.}(2009)\citenamefont {Jones},
  \citenamefont {Karlen}, \citenamefont {Fitzsimons}, \citenamefont {Ardavan},
  \citenamefont {Benjamin}, \citenamefont {Briggs},\ and\ \citenamefont
  {Morton}}]{JonesKarlenFitzsimons2009}%
  \BibitemOpen
  \bibfield  {author} {\bibinfo {author} {\bibfnamefont {J.~A.}\ \bibnamefont
  {Jones}}, \bibinfo {author} {\bibfnamefont {S.~D.}\ \bibnamefont {Karlen}},
  \bibinfo {author} {\bibfnamefont {J.}~\bibnamefont {Fitzsimons}}, \bibinfo
  {author} {\bibfnamefont {A.}~\bibnamefont {Ardavan}}, \bibinfo {author}
  {\bibfnamefont {S.~C.}\ \bibnamefont {Benjamin}}, \bibinfo {author}
  {\bibfnamefont {G.~A.~D.}\ \bibnamefont {Briggs}}, \ and\ \bibinfo {author}
  {\bibfnamefont {J.~J.~L.}\ \bibnamefont {Morton}},\ }\href@noop {} {\bibfield
   {journal} {\bibinfo  {journal} {Science}\ }\textbf {\bibinfo {volume}
  {324}},\ \bibinfo {pages} {1166} (\bibinfo {year} {2009})}\BibitemShut
  {NoStop}%
\bibitem [{\citenamefont {Zwierz}\ \emph {et~al.}(2012)\citenamefont {Zwierz},
  \citenamefont {P\'erez-Delgado},\ and\ \citenamefont
  {Kok}}]{ZwierzPerez-DelKok2012}%
  \BibitemOpen
  \bibfield  {author} {\bibinfo {author} {\bibfnamefont {M.}~\bibnamefont
  {Zwierz}}, \bibinfo {author} {\bibfnamefont {C.~A.}\ \bibnamefont
  {P\'erez-Delgado}}, \ and\ \bibinfo {author} {\bibfnamefont {P.}~\bibnamefont
  {Kok}},\ }\href@noop {} {\bibfield  {journal} {\bibinfo  {journal} {Phys.
  Rev. A}\ }\textbf {\bibinfo {volume} {85}},\ \bibinfo {pages} {042112}
  (\bibinfo {year} {2012})}\BibitemShut {NoStop}%
\bibitem [{\citenamefont {Aharonov}\ \emph {et~al.}(1988)\citenamefont
  {Aharonov}, \citenamefont {Albert},\ and\ \citenamefont
  {Vaidman}}]{AharonovAlbertVaidman1988}%
  \BibitemOpen
  \bibfield  {author} {\bibinfo {author} {\bibfnamefont {Y.}~\bibnamefont
  {Aharonov}}, \bibinfo {author} {\bibfnamefont {D.~Z.}\ \bibnamefont
  {Albert}}, \ and\ \bibinfo {author} {\bibfnamefont {L.}~\bibnamefont
  {Vaidman}},\ }\href@noop {} {\bibfield  {journal} {\bibinfo  {journal}
  {Physical Review Letters}\ }\textbf {\bibinfo {volume} {60}},\ \bibinfo
  {pages} {1351} (\bibinfo {year} {1988})}\BibitemShut {NoStop}%
\bibitem [{\citenamefont {Duck}\ \emph {et~al.}(1989)\citenamefont {Duck},
  \citenamefont {Stevenson},\ and\ \citenamefont
  {Sudarshan}}]{DuckStevensonSudarshan1989}%
  \BibitemOpen
  \bibfield  {author} {\bibinfo {author} {\bibfnamefont {I.~M.}\ \bibnamefont
  {Duck}}, \bibinfo {author} {\bibfnamefont {P.~M.}\ \bibnamefont {Stevenson}},
  \ and\ \bibinfo {author} {\bibfnamefont {E.~C.~G.}\ \bibnamefont
  {Sudarshan}},\ }\href@noop {} {\bibfield  {journal} {\bibinfo  {journal}
  {Phys. Rev. D}\ }\textbf {\bibinfo {volume} {40}},\ \bibinfo {pages} {2112}
  (\bibinfo {year} {1989})}\BibitemShut {NoStop}%
\bibitem [{\citenamefont {Wu}\ and\ \citenamefont {Li}(2011)}]{WuLi2011}%
  \BibitemOpen
  \bibfield  {author} {\bibinfo {author} {\bibfnamefont {S.}~\bibnamefont
  {Wu}}\ and\ \bibinfo {author} {\bibfnamefont {Y.}~\bibnamefont {Li}},\
  }\href@noop {} {\bibfield  {journal} {\bibinfo  {journal} {Phys. Rev. A}\
  }\textbf {\bibinfo {volume} {83}},\ \bibinfo {pages} {052106} (\bibinfo
  {year} {2011})}\BibitemShut {NoStop}%
\bibitem [{\citenamefont {Kofman}\ \emph {et~al.}(2012)\citenamefont {Kofman},
  \citenamefont {Ashhab},\ and\ \citenamefont {Nori}}]{KofmanAshhabNori2012}%
  \BibitemOpen
  \bibfield  {author} {\bibinfo {author} {\bibfnamefont {A.~G.}\ \bibnamefont
  {Kofman}}, \bibinfo {author} {\bibfnamefont {S.}~\bibnamefont {Ashhab}}, \
  and\ \bibinfo {author} {\bibfnamefont {F.}~\bibnamefont {Nori}},\ }\href@noop
  {} {\bibfield  {journal} {\bibinfo  {journal} {Physics Reports}\ }\textbf
  {\bibinfo {volume} {520}},\ \bibinfo {pages} {43 } (\bibinfo {year}
  {2012})}\BibitemShut {NoStop}%
\bibitem [{\citenamefont {Nakamura}\ \emph {et~al.}(2012)\citenamefont
  {Nakamura}, \citenamefont {Nishizawa},\ and\ \citenamefont
  {Fujimoto}}]{NakamuraNishizawaFujimoto2012}%
  \BibitemOpen
  \bibfield  {author} {\bibinfo {author} {\bibfnamefont {K.}~\bibnamefont
  {Nakamura}}, \bibinfo {author} {\bibfnamefont {A.}~\bibnamefont {Nishizawa}},
  \ and\ \bibinfo {author} {\bibfnamefont {M.-K.}\ \bibnamefont {Fujimoto}},\
  }\href@noop {} {\bibfield  {journal} {\bibinfo  {journal} {Phys. Rev. A}\
  }\textbf {\bibinfo {volume} {85}},\ \bibinfo {pages} {012113} (\bibinfo
  {year} {2012})}\BibitemShut {NoStop}%
\bibitem [{\citenamefont {Ritchie}\ \emph {et~al.}(1991)\citenamefont
  {Ritchie}, \citenamefont {Story},\ and\ \citenamefont
  {Hulet}}]{RitchieStoryHulet1991}%
  \BibitemOpen
  \bibfield  {author} {\bibinfo {author} {\bibfnamefont {N.~W.~M.}\
  \bibnamefont {Ritchie}}, \bibinfo {author} {\bibfnamefont {J.~G.}\
  \bibnamefont {Story}}, \ and\ \bibinfo {author} {\bibfnamefont {R.~G.}\
  \bibnamefont {Hulet}},\ }\href@noop {} {\bibfield  {journal} {\bibinfo
  {journal} {Phys. Rev. Lett.}\ }\textbf {\bibinfo {volume} {66}},\ \bibinfo
  {pages} {1107} (\bibinfo {year} {1991})}\BibitemShut {NoStop}%
\bibitem [{\citenamefont {Pryde}\ \emph {et~al.}(2005)\citenamefont {Pryde},
  \citenamefont {O'Brien}, \citenamefont {White}, \citenamefont {Ralph},\ and\
  \citenamefont {Wiseman}}]{PrydeOBrienWhite2005}%
  \BibitemOpen
  \bibfield  {author} {\bibinfo {author} {\bibfnamefont {G.~J.}\ \bibnamefont
  {Pryde}}, \bibinfo {author} {\bibfnamefont {J.~L.}\ \bibnamefont {O'Brien}},
  \bibinfo {author} {\bibfnamefont {A.~G.}\ \bibnamefont {White}}, \bibinfo
  {author} {\bibfnamefont {T.~C.}\ \bibnamefont {Ralph}}, \ and\ \bibinfo
  {author} {\bibfnamefont {H.~M.}\ \bibnamefont {Wiseman}},\ }\href@noop {}
  {\bibfield  {journal} {\bibinfo  {journal} {Phys. Rev. Lett.}\ }\textbf
  {\bibinfo {volume} {94}},\ \bibinfo {pages} {220405} (\bibinfo {year}
  {2005})}\BibitemShut {NoStop}%
\bibitem [{\citenamefont {Hosten}\ and\ \citenamefont
  {Kwiat}(2008)}]{HostenKwiat2008}%
  \BibitemOpen
  \bibfield  {author} {\bibinfo {author} {\bibfnamefont {O.}~\bibnamefont
  {Hosten}}\ and\ \bibinfo {author} {\bibfnamefont {P.}~\bibnamefont {Kwiat}},\
  }\href@noop {} {\bibfield  {journal} {\bibinfo  {journal} {Science}\ }\textbf
  {\bibinfo {volume} {319}},\ \bibinfo {pages} {787} (\bibinfo {year}
  {2008})}\BibitemShut {NoStop}%
\bibitem [{\citenamefont {Dixon}\ \emph {et~al.}(2009)\citenamefont {Dixon},
  \citenamefont {Starling}, \citenamefont {Jordan},\ and\ \citenamefont
  {Howell}}]{DixonStarlingJordan2009}%
  \BibitemOpen
  \bibfield  {author} {\bibinfo {author} {\bibfnamefont {P.~B.}\ \bibnamefont
  {Dixon}}, \bibinfo {author} {\bibfnamefont {D.~J.}\ \bibnamefont {Starling}},
  \bibinfo {author} {\bibfnamefont {A.~N.}\ \bibnamefont {Jordan}}, \ and\
  \bibinfo {author} {\bibfnamefont {J.~C.}\ \bibnamefont {Howell}},\
  }\href@noop {} {\bibfield  {journal} {\bibinfo  {journal} {Phys. Rev. Lett.}\
  }\textbf {\bibinfo {volume} {102}},\ \bibinfo {pages} {173601} (\bibinfo
  {year} {2009})}\BibitemShut {NoStop}%
\bibitem [{\citenamefont {Kedem}(2012)}]{Kedem2012}%
  \BibitemOpen
  \bibfield  {author} {\bibinfo {author} {\bibfnamefont {Y.}~\bibnamefont
  {Kedem}},\ }\href@noop {} {\bibfield  {journal} {\bibinfo  {journal} {Phys.
  Rev. A}\ }\textbf {\bibinfo {volume} {85}},\ \bibinfo {pages} {060102}
  (\bibinfo {year} {2012})}\BibitemShut {NoStop}%
\bibitem [{\citenamefont {Zhu}\ \emph {et~al.}(2011)\citenamefont {Zhu},
  \citenamefont {Zhang}, \citenamefont {Pang}, \citenamefont {Qiao},
  \citenamefont {Liu},\ and\ \citenamefont {Wu}}]{ZhuZhangPang2011}%
  \BibitemOpen
  \bibfield  {author} {\bibinfo {author} {\bibfnamefont {X.}~\bibnamefont
  {Zhu}}, \bibinfo {author} {\bibfnamefont {Y.}~\bibnamefont {Zhang}}, \bibinfo
  {author} {\bibfnamefont {S.}~\bibnamefont {Pang}}, \bibinfo {author}
  {\bibfnamefont {C.}~\bibnamefont {Qiao}}, \bibinfo {author} {\bibfnamefont
  {Q.}~\bibnamefont {Liu}}, \ and\ \bibinfo {author} {\bibfnamefont
  {S.}~\bibnamefont {Wu}},\ }\href@noop {} {\bibfield  {journal} {\bibinfo
  {journal} {Phys. Rev. A}\ }\textbf {\bibinfo {volume} {84}},\ \bibinfo
  {pages} {052111} (\bibinfo {year} {2011})}\BibitemShut {NoStop}%
\bibitem [{\citenamefont {Starling}\ \emph {et~al.}(2009)\citenamefont
  {Starling}, \citenamefont {Dixon}, \citenamefont {Jordan},\ and\
  \citenamefont {Howell}}]{StarlingDixonJordan2009}%
  \BibitemOpen
  \bibfield  {author} {\bibinfo {author} {\bibfnamefont {D.~J.}\ \bibnamefont
  {Starling}}, \bibinfo {author} {\bibfnamefont {P.~B.}\ \bibnamefont {Dixon}},
  \bibinfo {author} {\bibfnamefont {A.~N.}\ \bibnamefont {Jordan}}, \ and\
  \bibinfo {author} {\bibfnamefont {J.~C.}\ \bibnamefont {Howell}},\
  }\href@noop {} {\bibfield  {journal} {\bibinfo  {journal} {Phys. Rev. A}\
  }\textbf {\bibinfo {volume} {80}},\ \bibinfo {pages} {041803} (\bibinfo
  {year} {2009})}\BibitemShut {NoStop}%
\bibitem [{\citenamefont {Schlosshauer}(2005)}]{Schlosshau2005}%
  \BibitemOpen
  \bibfield  {author} {\bibinfo {author} {\bibfnamefont {M.}~\bibnamefont
  {Schlosshauer}},\ }\href@noop {} {\bibfield  {journal} {\bibinfo  {journal}
  {Rev. Mod. Phys.}\ }\textbf {\bibinfo {volume} {76}},\ \bibinfo {pages}
  {1267} (\bibinfo {year} {2005})}\BibitemShut {NoStop}%
\bibitem [{\citenamefont {Shaji}\ and\ \citenamefont
  {Caves}(2007)}]{ShajiCaves2007}%
  \BibitemOpen
  \bibfield  {author} {\bibinfo {author} {\bibfnamefont {A.}~\bibnamefont
  {Shaji}}\ and\ \bibinfo {author} {\bibfnamefont {C.~M.}\ \bibnamefont
  {Caves}},\ }\href@noop {} {\bibfield  {journal} {\bibinfo  {journal} {Phys.
  Rev. A}\ }\textbf {\bibinfo {volume} {76}},\ \bibinfo {pages} {032111}
  (\bibinfo {year} {2007})}\BibitemShut {NoStop}%
\bibitem [{\citenamefont {Escher}\ \emph {et~al.}(2011)\citenamefont {Escher},
  \citenamefont {de~Matos~Filho},\ and\ \citenamefont
  {Davidovich}}]{EscherMatos-FilDavidovic2011}%
  \BibitemOpen
  \bibfield  {author} {\bibinfo {author} {\bibfnamefont {B.~M.}\ \bibnamefont
  {Escher}}, \bibinfo {author} {\bibfnamefont {R.~L.}\ \bibnamefont
  {de~Matos~Filho}}, \ and\ \bibinfo {author} {\bibfnamefont {L.}~\bibnamefont
  {Davidovich}},\ }\href@noop {} {\bibfield  {journal} {\bibinfo  {journal}
  {Nat Phys}\ }\textbf {\bibinfo {volume} {7}},\ \bibinfo {pages} {406}
  (\bibinfo {year} {2011})}\BibitemShut {NoStop}%
\bibitem [{\citenamefont {Hofmann}\ \emph {et~al.}(2012)\citenamefont
  {Hofmann}, \citenamefont {Goggin}, \citenamefont {Almeida},\ and\
  \citenamefont {Barbieri}}]{HofmannGogginAlmeida2012}%
  \BibitemOpen
  \bibfield  {author} {\bibinfo {author} {\bibfnamefont {H.~F.}\ \bibnamefont
  {Hofmann}}, \bibinfo {author} {\bibfnamefont {M.~E.}\ \bibnamefont {Goggin}},
  \bibinfo {author} {\bibfnamefont {M.~P.}\ \bibnamefont {Almeida}}, \ and\
  \bibinfo {author} {\bibfnamefont {M.}~\bibnamefont {Barbieri}},\ }\href@noop
  {} {\bibfield  {journal} {\bibinfo  {journal} {Phys. Rev. A}\ }\textbf
  {\bibinfo {volume} {86}},\ \bibinfo {pages} {040102} (\bibinfo {year}
  {2012})}\BibitemShut {NoStop}%
\bibitem [{\citenamefont {Burkard}(2009)}]{Burkard2009}%
  \BibitemOpen
  \bibfield  {author} {\bibinfo {author} {\bibfnamefont {G.}~\bibnamefont
  {Burkard}},\ }\href@noop {} {\bibfield  {journal} {\bibinfo  {journal} {Phys.
  Rev. B}\ }\textbf {\bibinfo {volume} {79}},\ \bibinfo {pages} {125317}
  (\bibinfo {year} {2009})}\BibitemShut {NoStop}%
\bibitem [{\citenamefont {Parks}\ and\ \citenamefont
  {Gray}(2011)}]{ParksGray2011}%
  \BibitemOpen
  \bibfield  {author} {\bibinfo {author} {\bibfnamefont {A.~D.}\ \bibnamefont
  {Parks}}\ and\ \bibinfo {author} {\bibfnamefont {J.~E.}\ \bibnamefont
  {Gray}},\ }\href@noop {} {\bibfield  {journal} {\bibinfo  {journal} {Phys.
  Rev. A}\ }\textbf {\bibinfo {volume} {84}},\ \bibinfo {pages} {012116}
  (\bibinfo {year} {2011})}\BibitemShut {NoStop}%
\bibitem [{\citenamefont {Ashhab}\ \emph {et~al.}(2009)\citenamefont {Ashhab},
  \citenamefont {You},\ and\ \citenamefont {Nori}}]{AshhabYouNori2009}%
  \BibitemOpen
  \bibfield  {author} {\bibinfo {author} {\bibfnamefont {S.}~\bibnamefont
  {Ashhab}}, \bibinfo {author} {\bibfnamefont {J.~Q.}\ \bibnamefont {You}}, \
  and\ \bibinfo {author} {\bibfnamefont {F.}~\bibnamefont {Nori}},\ }\href@noop
  {} {\bibfield  {journal} {\bibinfo  {journal} {New Journal of Physics}\
  }\textbf {\bibinfo {volume} {11}},\ \bibinfo {pages} {083017} (\bibinfo
  {year} {2009})}\BibitemShut {NoStop}%
\bibitem [{\citenamefont {Nishizawa}\ \emph {et~al.}(2012)\citenamefont
  {Nishizawa}, \citenamefont {Nakamura},\ and\ \citenamefont
  {Fujimoto}}]{NishizawaNakamuraFujimoto2012}%
  \BibitemOpen
  \bibfield  {author} {\bibinfo {author} {\bibfnamefont {A.}~\bibnamefont
  {Nishizawa}}, \bibinfo {author} {\bibfnamefont {K.}~\bibnamefont {Nakamura}},
  \ and\ \bibinfo {author} {\bibfnamefont {M.-K.}\ \bibnamefont {Fujimoto}},\
  }\href@noop {} {\bibfield  {journal} {\bibinfo  {journal} {Phys. Rev. A}\
  }\textbf {\bibinfo {volume} {85}},\ \bibinfo {pages} {062108} (\bibinfo
  {year} {2012})}\BibitemShut {NoStop}%
\bibitem [{\citenamefont {Brunner}\ and\ \citenamefont
  {Simon}(2010)}]{BrunnerSimon2010}%
  \BibitemOpen
  \bibfield  {author} {\bibinfo {author} {\bibfnamefont {N.}~\bibnamefont
  {Brunner}}\ and\ \bibinfo {author} {\bibfnamefont {C.}~\bibnamefont
  {Simon}},\ }\href@noop {} {\bibfield  {journal} {\bibinfo  {journal} {Phys.
  Rev. Lett.}\ }\textbf {\bibinfo {volume} {105}},\ \bibinfo {pages} {010405}
  (\bibinfo {year} {2010})}\BibitemShut {NoStop}%
\bibitem [{\citenamefont {Feizpour}\ \emph {et~al.}(2011)\citenamefont
  {Feizpour}, \citenamefont {Xing},\ and\ \citenamefont
  {Steinberg}}]{FeizpourXingSteinberg2011}%
  \BibitemOpen
  \bibfield  {author} {\bibinfo {author} {\bibfnamefont {A.}~\bibnamefont
  {Feizpour}}, \bibinfo {author} {\bibfnamefont {X.}~\bibnamefont {Xing}}, \
  and\ \bibinfo {author} {\bibfnamefont {A.~M.}\ \bibnamefont {Steinberg}},\
  }\href@noop {} {\bibfield  {journal} {\bibinfo  {journal} {Phys. Rev. Lett.}\
  }\textbf {\bibinfo {volume} {107}},\ \bibinfo {pages} {133603} (\bibinfo
  {year} {2011})}\BibitemShut {NoStop}%
\bibitem [{\citenamefont {Braunstein}\ and\ \citenamefont
  {Caves}(1994)}]{BraunsteinCaves1994}%
  \BibitemOpen
  \bibfield  {author} {\bibinfo {author} {\bibfnamefont {S.~L.}\ \bibnamefont
  {Braunstein}}\ and\ \bibinfo {author} {\bibfnamefont {C.~M.}\ \bibnamefont
  {Caves}},\ }\href@noop {} {\bibfield  {journal} {\bibinfo  {journal} {Phys.
  Rev. Lett.}\ }\textbf {\bibinfo {volume} {72}},\ \bibinfo {pages} {3439}
  (\bibinfo {year} {1994})}\BibitemShut {NoStop}%
\bibitem [{\citenamefont {Helstrom}(1969)}]{Helstrom1969}%
  \BibitemOpen
  \bibfield  {author} {\bibinfo {author} {\bibfnamefont {C.~W.}\ \bibnamefont
  {Helstrom}},\ }\href@noop {} {\bibfield  {journal} {\bibinfo  {journal}
  {Journal of Statistical Physics}\ }\textbf {\bibinfo {volume} {1}},\ \bibinfo
  {pages} {231} (\bibinfo {year} {1969})}\BibitemShut {NoStop}%
\bibitem [{\citenamefont {Kholevo}(1982)}]{Kholevo1982}%
  \BibitemOpen
  \bibfield  {author} {\bibinfo {author} {\bibfnamefont {A.~S.}\ \bibnamefont
  {Kholevo}},\ }\href@noop {} {\emph {\bibinfo {title} {Probabilistic and
  statistical aspects of quantum theory}}}\ (\bibinfo  {publisher}
  {North-Holland Pub. Co. ;},\ \bibinfo {address} {Amsterdam ;},\ \bibinfo
  {year} {1982})\BibitemShut {NoStop}%
\bibitem [{\citenamefont {Fisher}(1925)}]{Fisher1925}%
  \BibitemOpen
  \bibfield  {author} {\bibinfo {author} {\bibfnamefont {R.~A.}\ \bibnamefont
  {Fisher}},\ }\href@noop {} {\bibfield  {journal} {\bibinfo  {journal}
  {Mathematical Proceedings of the Cambridge Philosophical Society}\ }\textbf
  {\bibinfo {volume} {22}},\ \bibinfo {pages} {700} (\bibinfo {year}
  {1925})}\BibitemShut {NoStop}%
\bibitem [{\citenamefont {Hofmann}(2011)}]{Hofmann2011}%
  \BibitemOpen
  \bibfield  {author} {\bibinfo {author} {\bibfnamefont {H.~F.}\ \bibnamefont
  {Hofmann}},\ }\href@noop {} {\bibfield  {journal} {\bibinfo  {journal} {Phys.
  Rev. A}\ }\textbf {\bibinfo {volume} {83}},\ \bibinfo {pages} {022106}
  (\bibinfo {year} {2011})}\BibitemShut {NoStop}%
\bibitem [{\citenamefont {Myung}(2003)}]{Myung2003}%
  \BibitemOpen
  \bibfield  {author} {\bibinfo {author} {\bibfnamefont {I.~J.}\ \bibnamefont
  {Myung}},\ }\href@noop {} {\bibfield  {journal} {\bibinfo  {journal} {J.
  Math. Psychol.}\ }\textbf {\bibinfo {volume} {47}},\ \bibinfo {pages} {90}
  (\bibinfo {year} {2003})}\BibitemShut {NoStop}%
\bibitem [{\citenamefont {Brivio}\ \emph {et~al.}(2010)\citenamefont {Brivio},
  \citenamefont {Cialdi}, \citenamefont {Vezzoli}, \citenamefont {Gebrehiwot},
  \citenamefont {Genoni}, \citenamefont {Olivares},\ and\ \citenamefont
  {Paris}}]{BrivioCialdiVezzoli2010}%
  \BibitemOpen
  \bibfield  {author} {\bibinfo {author} {\bibfnamefont {D.}~\bibnamefont
  {Brivio}}, \bibinfo {author} {\bibfnamefont {S.}~\bibnamefont {Cialdi}},
  \bibinfo {author} {\bibfnamefont {S.}~\bibnamefont {Vezzoli}}, \bibinfo
  {author} {\bibfnamefont {B.~T.}\ \bibnamefont {Gebrehiwot}}, \bibinfo
  {author} {\bibfnamefont {M.~G.}\ \bibnamefont {Genoni}}, \bibinfo {author}
  {\bibfnamefont {S.}~\bibnamefont {Olivares}}, \ and\ \bibinfo {author}
  {\bibfnamefont {M.~G.~A.}\ \bibnamefont {Paris}},\ }\href@noop {} {\bibfield
  {journal} {\bibinfo  {journal} {Phys. Rev. A}\ }\textbf {\bibinfo {volume}
  {81}},\ \bibinfo {pages} {012305} (\bibinfo {year} {2010})}\BibitemShut
  {NoStop}%
\bibitem [{\citenamefont {Paris}(2009)}]{Paris2009}%
  \BibitemOpen
  \bibfield  {author} {\bibinfo {author} {\bibfnamefont {M.~G.~A.}\
  \bibnamefont {Paris}},\ }\href@noop {} {\bibfield  {journal} {\bibinfo
  {journal} {Int. J. Quant. Inf.}\ }\textbf {\bibinfo {volume} {7}},\ \bibinfo
  {pages} {125} (\bibinfo {year} {2009})}\BibitemShut {NoStop}%
\bibitem [{\citenamefont {Cramer}(1946)}]{Cramer1999}%
  \BibitemOpen
  \bibfield  {author} {\bibinfo {author} {\bibfnamefont {H.}~\bibnamefont
  {Cramer}},\ }\href@noop {} {\emph {\bibinfo {title} {Mathematical methods of
  statistics}}}\ (\bibinfo  {publisher} {Princeton University Press},\ \bibinfo
  {address} {Princeton},\ \bibinfo {year} {1946})\BibitemShut {NoStop}%
\bibitem [{\citenamefont {Modi}\ \emph {et~al.}(2011)\citenamefont {Modi},
  \citenamefont {Cable}, \citenamefont {Williamson},\ and\ \citenamefont
  {Vedral}}]{ModiCableWilliamson2011}%
  \BibitemOpen
  \bibfield  {author} {\bibinfo {author} {\bibfnamefont {K.}~\bibnamefont
  {Modi}}, \bibinfo {author} {\bibfnamefont {H.}~\bibnamefont {Cable}},
  \bibinfo {author} {\bibfnamefont {M.}~\bibnamefont {Williamson}}, \ and\
  \bibinfo {author} {\bibfnamefont {V.}~\bibnamefont {Vedral}},\ }\href@noop {}
  {\bibfield  {journal} {\bibinfo  {journal} {Phys. Rev. X}\ }\textbf {\bibinfo
  {volume} {1}},\ \bibinfo {pages} {021022} (\bibinfo {year}
  {2011})}\BibitemShut {NoStop}%
\bibitem [{\citenamefont {Schaffry}\ \emph {et~al.}(2010)\citenamefont
  {Schaffry}, \citenamefont {Gauger}, \citenamefont {Morton}, \citenamefont
  {Fitzsimons}, \citenamefont {Benjamin},\ and\ \citenamefont
  {Lovett}}]{SchaffryGaugerMorton2010}%
  \BibitemOpen
  \bibfield  {author} {\bibinfo {author} {\bibfnamefont {M.}~\bibnamefont
  {Schaffry}}, \bibinfo {author} {\bibfnamefont {E.~M.}\ \bibnamefont
  {Gauger}}, \bibinfo {author} {\bibfnamefont {J.~J.~L.}\ \bibnamefont
  {Morton}}, \bibinfo {author} {\bibfnamefont {J.}~\bibnamefont {Fitzsimons}},
  \bibinfo {author} {\bibfnamefont {S.~C.}\ \bibnamefont {Benjamin}}, \ and\
  \bibinfo {author} {\bibfnamefont {B.~W.}\ \bibnamefont {Lovett}},\
  }\href@noop {} {\bibfield  {journal} {\bibinfo  {journal} {Phys. Rev. A}\
  }\textbf {\bibinfo {volume} {82}},\ \bibinfo {pages} {042114} (\bibinfo
  {year} {2010})}\BibitemShut {NoStop}%
\bibitem [{\citenamefont {Schaffry}\ \emph {et~al.}(2011)\citenamefont
  {Schaffry}, \citenamefont {Gauger}, \citenamefont {Morton},\ and\
  \citenamefont {Benjamin}}]{SchaffryGaugerMorton2011}%
  \BibitemOpen
  \bibfield  {author} {\bibinfo {author} {\bibfnamefont {M.}~\bibnamefont
  {Schaffry}}, \bibinfo {author} {\bibfnamefont {E.~M.}\ \bibnamefont
  {Gauger}}, \bibinfo {author} {\bibfnamefont {J.~J.~L.}\ \bibnamefont
  {Morton}}, \ and\ \bibinfo {author} {\bibfnamefont {S.~C.}\ \bibnamefont
  {Benjamin}},\ }\href@noop {} {\bibfield  {journal} {\bibinfo  {journal}
  {Phys. Rev. Lett.}\ }\textbf {\bibinfo {volume} {107}},\ \bibinfo {pages}
  {207210} (\bibinfo {year} {2011})}\BibitemShut {NoStop}%
\bibitem [{Note1()}]{Note1}%
  \BibitemOpen
  \bibinfo {note} {Best results are obtained when the ancilla measurement basis
  is unbiased with respect to the axis about which the ancilla
  rotates}\BibitemShut {NoStop}%
\bibitem [{Note2()}]{Note2}%
  \BibitemOpen
  \bibinfo {note} {Identity matrices are implied in the following expression,
  i.e.~${\delimiter "426830A \psi _f|_s=\delimiter "426830A \psi _f|\otimes
  \protect \mathbb {I}}$ etc.}\BibitemShut {Stop}%
\bibitem [{\citenamefont {Teklu}\ \emph {et~al.}(2010)\citenamefont {Teklu},
  \citenamefont {Genoni}, \citenamefont {Olivares},\ and\ \citenamefont
  {Paris}}]{TekluGenoniOlivares2010}%
  \BibitemOpen
  \bibfield  {author} {\bibinfo {author} {\bibfnamefont {B.}~\bibnamefont
  {Teklu}}, \bibinfo {author} {\bibfnamefont {M.~G.}\ \bibnamefont {Genoni}},
  \bibinfo {author} {\bibfnamefont {S.}~\bibnamefont {Olivares}}, \ and\
  \bibinfo {author} {\bibfnamefont {M.~G.~A.}\ \bibnamefont {Paris}},\
  }\href@noop {} {\bibfield  {journal} {\bibinfo  {journal} {Physi. Scr.}\
  }\textbf {\bibinfo {volume} {2010}},\ \bibinfo {pages} {014062} (\bibinfo
  {year} {2010})}\BibitemShut {NoStop}%
\bibitem [{Note3()}]{Note3}%
  \BibitemOpen
  \bibinfo {note} {One can choose instead to have the preselection unbiased
  w.r.t.~the control observable, as for the $H_{\protect \textrm {anc}}$ above,
  but one does no better in that case.}\BibitemShut {Stop}%
\bibitem [{\citenamefont {Zilberberg}\ \emph {et~al.}(2011)\citenamefont
  {Zilberberg}, \citenamefont {Romito},\ and\ \citenamefont
  {Gefen}}]{ZilberbergRomitoGefen2011}%
  \BibitemOpen
  \bibfield  {author} {\bibinfo {author} {\bibfnamefont {O.}~\bibnamefont
  {Zilberberg}}, \bibinfo {author} {\bibfnamefont {A.}~\bibnamefont {Romito}},
  \ and\ \bibinfo {author} {\bibfnamefont {Y.}~\bibnamefont {Gefen}},\
  }\href@noop {} {\bibfield  {journal} {\bibinfo  {journal} {Phys. Rev. Lett.}\
  }\textbf {\bibinfo {volume} {106}},\ \bibinfo {pages} {080405} (\bibinfo
  {year} {2011})}\BibitemShut {NoStop}%
\bibitem [{\citenamefont {Shikano}\ and\ \citenamefont
  {Tanaka}(2011)}]{ShikanoTanaka2011}%
  \BibitemOpen
  \bibfield  {author} {\bibinfo {author} {\bibfnamefont {Y.}~\bibnamefont
  {Shikano}}\ and\ \bibinfo {author} {\bibfnamefont {S.}~\bibnamefont
  {Tanaka}},\ }\href@noop {} {\bibfield  {journal} {\bibinfo  {journal} {EPL
  (Europhysics Letters)}\ }\textbf {\bibinfo {volume} {96}},\ \bibinfo {pages}
  {40002} (\bibinfo {year} {2011})}\BibitemShut {NoStop}%
\bibitem [{\citenamefont {Wu}\ and\ \citenamefont
  {Molmer}(2009)}]{WuMolmer2009}%
  \BibitemOpen
  \bibfield  {author} {\bibinfo {author} {\bibfnamefont {S.}~\bibnamefont
  {Wu}}\ and\ \bibinfo {author} {\bibfnamefont {K.}~\bibnamefont {Molmer}},\
  }\href@noop {} {\bibfield  {journal} {\bibinfo  {journal} {Physics Letters
  A}\ }\textbf {\bibinfo {volume} {374}},\ \bibinfo {pages} {34 } (\bibinfo
  {year} {2009})}\BibitemShut {NoStop}%
\bibitem [{\citenamefont {Cho}\ \emph {et~al.}(2010)\citenamefont {Cho},
  \citenamefont {Lim}, \citenamefont {Ra},\ and\ \citenamefont
  {Kim}}]{ChoLimRa2010}%
  \BibitemOpen
  \bibfield  {author} {\bibinfo {author} {\bibfnamefont {Y.-W.}\ \bibnamefont
  {Cho}}, \bibinfo {author} {\bibfnamefont {H.-T.}\ \bibnamefont {Lim}},
  \bibinfo {author} {\bibfnamefont {Y.-S.}\ \bibnamefont {Ra}}, \ and\ \bibinfo
  {author} {\bibfnamefont {Y.-H.}\ \bibnamefont {Kim}},\ }\href@noop {}
  {\bibfield  {journal} {\bibinfo  {journal} {New Journal of Physics}\ }\textbf
  {\bibinfo {volume} {12}},\ \bibinfo {pages} {023036} (\bibinfo {year}
  {2010})}\BibitemShut {NoStop}%
\bibitem [{\citenamefont {Shikano}\ and\ \citenamefont
  {Hosoya}(2010)}]{ShikanoHosoya2010}%
  \BibitemOpen
  \bibfield  {author} {\bibinfo {author} {\bibfnamefont {Y.}~\bibnamefont
  {Shikano}}\ and\ \bibinfo {author} {\bibfnamefont {A.}~\bibnamefont
  {Hosoya}},\ }\href@noop {} {\bibfield  {journal} {\bibinfo  {journal}
  {Journal of Physics A: Mathematical and Theoretical}\ }\textbf {\bibinfo
  {volume} {43}},\ \bibinfo {pages} {025304} (\bibinfo {year}
  {2010})}\BibitemShut {NoStop}%
\bibitem [{\citenamefont {Thomas}\ and\ \citenamefont
  {Romito}(2012)}]{ThomasRomito2012}%
  \BibitemOpen
  \bibfield  {author} {\bibinfo {author} {\bibfnamefont {M.}~\bibnamefont
  {Thomas}}\ and\ \bibinfo {author} {\bibfnamefont {A.}~\bibnamefont
  {Romito}},\ }\href@noop {} {\bibfield  {journal} {\bibinfo  {journal} {Phys.
  Rev. B}\ }\textbf {\bibinfo {volume} {86}},\ \bibinfo {pages} {235419}
  (\bibinfo {year} {2012})}\BibitemShut {NoStop}%
\bibitem [{\citenamefont {Steinberg}(2010)}]{Steinberg2010}%
  \BibitemOpen
  \bibfield  {author} {\bibinfo {author} {\bibfnamefont {A.~M.}\ \bibnamefont
  {Steinberg}},\ }\href@noop {} {\bibfield  {journal} {\bibinfo  {journal}
  {Nature}\ }\textbf {\bibinfo {volume} {463}},\ \bibinfo {pages} {890}
  (\bibinfo {year} {2010})}\BibitemShut {NoStop}%
\end{thebibliography}%
\end{document}